\documentclass[aps,showpacs,prb,floatfix,twocolumn,citeautoscript]{revtex4-1}
\usepackage{graphicx}
\usepackage{amsmath}
\usepackage{amssymb}
\usepackage{bm}
\usepackage{dcolumn}
\usepackage{subfigure}
\usepackage{units}
\usepackage{hyperref}
\usepackage[usenames]{color}
\usepackage{booktabs}
\usepackage{multirow}
\usepackage{siunitx}
\usepackage{wrapfig}
\hypersetup{pdfborder=0 0 0,colorlinks=true,citecolor=blue,linkcolor=blue}
\input epsf

\newcommand{\BN}{\textit{h}-BN}

\newcommand{\MS}{MoS$_2$}

\begin{document}

\title{A first-principles study of van der Waals interactions and lattice mismatch at \MS/metal interfaces}
\author{Mojtaba Farmanbar}
\email{m.farmanbar@utwente.nl}
\author{Geert Brocks}
\email{g.h.l.a.brocks@utwente.nl}
\affiliation{Faculty of Science and Technology and MESA$^+$ Institute for Nanotechnology, University of Twente, P.O. Box 217, 7500 AE Enschede, The Netherlands}

\begin{abstract}
We explore the adsorption of \MS\ on a range of metal substrates by means of first-principles density functional theory calculations. Including van der Waals forces in the density functional is essential to capture the interaction between \MS\ and a metal surface, and obtain reliable interface potential steps and Schottky barriers. Special care is taken to construct interface structures that have a mismatch between the \MS\ and the metal lattices of $<$1\%. \MS\ is chemisorbed on the early transition metal Ti, which leads to a strong perturbation of its (electronic) structure and a pinning of the Fermi level 0.54 eV below the \MS\ conduction band due to interface states. \MS\ is physisorbed on Au, where the bonding hardly perturbs the electronic structure. The bonding of \MS\ on other metals lies between these two extreme cases, with interface interactions for the late 3d transition metals Co, Ni, Cu and the simple metal Mg that are somewhat stronger than for the late 4d/5d transition metals Pd, Ag, Pt and the simple metal Al. Even a weak interaction, such as in the case of Al, gives interface states, however, with energies inside the \MS\ band gap, which pin the Fermi level below the conduction band.   
\end{abstract}

\date{\today}
\maketitle

\maketitle
\section{Introduction}

Transition metal dichalcogenides (TMDs) such as molybdenum disulfide (\MS) have layered structures, where the atoms within a TMD monolayer form a covalently bonded planar network, and the interaction between these layers is a weak, van der Waals interaction.\cite{Chhowalla:natchem13,Xu:chemrev13}  A monolayer of \MS\ consists of a layer of molybdenum atoms sandwiched between two layers of sulfur atoms. \MS\ monolayers can be exfoliated through micro-mechanical cleavage, similar to graphene or boron nitride.\cite{Geim:nat13} Unlike graphene (a metal), or boron nitride (an insulator), \MS\ is a semiconductor. Moreover, whereas bulk \MS\ has an indirect band gap (1.2 eV), monolayer \MS\ has a direct band gap ($\sim$1.8-1.9 eV), and shows a strong optical absorption and luminescence.\cite{Mak:prl10,Wang:nnano12} At present \MS, and TMDs in general, are vehemently pursued as promising materials for applications in electronics and optoelectronics.\cite{Wang:nnano12,Lembke:acchem15}

Contacting \MS\ to metal electrodes proves to be a problem; it tends to produce unexpectedly high interface resistances, indicative of a high Schottky barrier at the interface.\cite{Liu:acsnano12,Das:nanol13,ChenJR:nanol13,Fontana:scirep13,Kaushik:apl14,Kang:apl14} A high barrier could be caused by strong interface bonding creating interface states that pin the Fermi level,\cite{Farmanbar:prb15} or by weak bonding creating a potential step due to Pauli repulsion at the interface.\cite{Bokdam:prb14a,Bokdam:prb14b} The nature of the interaction at the \MS/metal interface is far from trivial. On the one hand, one could argue that, as \MS\ has no dangling bonds at its surface, its interaction with metal substrates should be weak and van-der-Waals-like. On the other hand, many metal species form (di)chalcogenide compounds,\cite{Mak:prl10,Wang:nnano12,ataca:jpcc12,Wissam:jcp141} and when adsorbing \MS\ onto a metal substrate, there could be a competition between the metal surface and the Mo atoms for interacting with the sulfur atoms at the interface. In that case, not only the \MS/metal bonding would be a much stronger chemical bonding, but also the structure and electronic structure of the \MS\ adsorbate could be significantly perturbed. 

In this paper we explore the adsorption of \MS\ on a variety of metal substrates by means of first-principles density functional theory (DFT) calculations, following up on work briefly reported in a short paper.\cite{Farmanbar:prb15} Previous DFT studies have concentrated foremost on the Schottky barrier formed at \MS/metal interfaces using the local density approximation (LDA).\cite{Perdew:prb81,Popov:prl12,Chen:nanol13,Kang:prx14,Gong:nanol14,Li:acs23} LDA gives a reasonable description of the adsorption of graphene and \BN\ on metal surfaces, but such results cannot be generalized to other systems, as it is known that LDA often leads to an unrealistic overbinding.\cite{Giovannetti:prl08,Laskowski:prb08,Khomyakov:prb11,Stradi:prl11,Olsen:prl11,Andersen:prb12,Janthon:jcp13,Bokdam:prb14a,Bokdam:prb14b} 
Other studies have used a generalized gradient approximation (GGA) functional, such as PBE,\cite{Perdew:prl96} which apparently works well for TMDs adsorbed on metals,\cite{Cakir:jmcc46,Cakir:prb89,Farmanbar:prb15} although it generally gives bad results for weakly bonded systems.\cite{Sachs:prb84,Hazrati:prb14}

Here we focus on the interface interaction and its implications for the structure and electronic structure of the \MS\ adsorbate and the Schottky barrier. We choose a wide range of metal substrates: the (111) surfaces of Al, Ni, Cu, Pd, Ag, Pt and Au, and the (0001) surfaces of Mg, Ti, and Co, which are expected to have a wide range of interaction strength with the adsorbate. As the interface interaction can vary from weak (physisorption) to strong (chemisorption), it is a priori not clear which DFT functional describes such bonding. We test and compare results obtained with a van der Waals functional, designed to describe weak, van der Waals, interactions,\cite{Dion:prl04,Thonhauser:prb76,Klimes:prb11} to results obtained with GGA and LDA functionals, which are conventionally used to describe chemical bonding. We assess the importance of van der Waals interactions for the interface interaction, and evaluate its effect on the structure and electronic structure of the \MS\ adsorbant. 

We consider the situation where a \MS\ layer is adsorbed as a whole on a metal substrate, making it more likely that the integrity of the \MS\ layer is preserved in the adsorption process. If the \MS/metal interaction is not too strong, and the \MS\ and metal surface lattices are not matched, the interface structure is likely to be incommensurable. In a supercell calculation one is forced to approximate such a structure by a commensurable one. Previous calculations have used small supercells, where in some cases appreciable artificial strain is generated because of the mismatch between the \MS\ and the metal surface lattices.\cite{Popov:prl12,Kang:prx14,Gong:nanol14,Li:acs23} We apply a strategy for choosing supercells such that the artificial strain is minimal, and test the influence of strain on the electronic properties of the interface. 

This paper is organized as follows. Section~\ref{sec:calculations} describes the DFT calculations, comparing different functionals in Sec.~\ref{sec:functionals} and discussing the effect of lattice mismatch in Sec.~\ref{sec:mismatch}. Results are discussed in Sec.~\ref{sec:results}, with the metal/\MS\ interaction in Sec.~\ref{sec:bonding} and its effects on the interface potential step and the Schottky barrier in Sec.~\ref{sec:sb}. Strong chemisorption is discussed in more detail in Sec.~\ref{sec:Ti}, and a summary and the conclusions are presented in Sec.~\ref{sec:conclusions}.

\maketitle
\section{Calculations}
\label{sec:calculations}
\subsection{Computational Methods}
We calculate ground-state energies and optimize geometries at the density functional theory (DFT) level, using projector-augmented waves (PAWs) as implemented in the VASP code.\cite{Kresse:prb93,Blochl:prb94b,Kresse:prb96,Kresse:prb99} The plane-wave kinetic-energy cutoff is set at 400 eV. The surface Brillouin zone is integrated with the Methfessel-Paxton technique using a smearing parameter of 0.05 eV,\cite{Methfessel:prb89} and a $k$-point sampling grid with a spacing of 0.01 \AA$^{-1}$. The \MS/metal interface is modeled as a slab of 4-6 layers of metal atoms with one or two layers of \MS\ adsorbed on one side and a vacuum region of $\sim$12 \AA. The in-plane supercell is chosen such as to minimize the mismatch between the \MS\ and metal lattices, which is discussed in more detail in Sec.~\ref{sec:mismatch}. A dipole correction is applied to avoid spurious interactions between periodic images of the slab.\cite{Neugebauer:prb46} We allow the positions of the atoms to relax until the force on each atom is smaller than 0.01 eV\AA$^{-1}$, except for the bottom layer of metal atoms, whose positions are kept fixed. The electronic self-consistency criterion is set to 10$^{-5}$ eV. 

It is well known that commonly used DFT exchange-correlation functionals, based upon LDA\cite{Perdew:prb81} or GGA,\cite{Perdew:prl96} give decent descriptions of covalent and ionic bonding, but they may fail for weakly bonded systems, as such functionals do not contain a description of van der Waals interactions. For example, GGA functionals such as PW91 or PBE,\cite{Perdew:prl96} do not capture the bonding between \BN\ or graphene layers, nor that between \BN\ or graphene and transition metal(111) surfaces.\cite{Giovan:prb76,Hazrati:prb14} A priori we don't know how important van der Waals interactions are in the bonding between \MS\ and a metal surface.  In Sec.~\ref{sec:functionals} we compare results obtained using a van der Waals density functional (vdW-DF),\cite{Dion:prl04,Thonhauser:prb76,Klimes:prb11} with results obtained with GGA and LDA functionals.

\begin{figure}[tb]
\includegraphics[width=0.7\columnwidth]{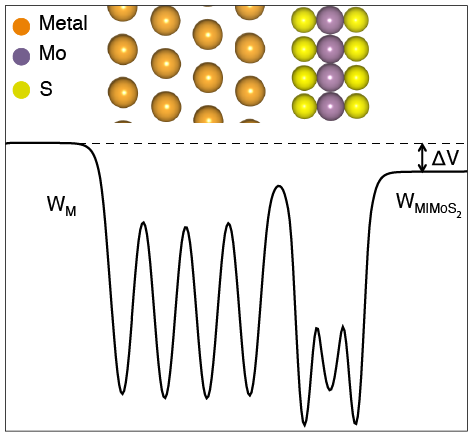}
\caption{(Color online) Side view of metal/\MS\ structure with corresponding plane-averaged electrostatic potential $\overline{V}(z)$. $\Delta{V}$ is defined as the difference between the work function on the metal side and on the \MS\ side. }
 \label{fig:deltaV}
\end{figure}

One way of visualizing bonding at a \MS/metal interface is by the electron density difference 
\begin{equation}
\Delta{n(\mathbf{r})} = n_\mathrm{ M \mid MoS_{2}}(\mathbf{r})  - n_\mathrm{M}(\mathbf{r}) - n_\mathrm{MoS_{2}}(\mathbf{r}), 
\label{eq:sb1}
\end{equation} 
where $n_\mathrm{ M \mid MoS_{2}}(\textbf{r})$, $n_\mathrm{M}(\textbf{r})$, and $n_\mathrm{MoS_{2}}(\textbf{r})$ are the electron densities of \MS\ adsorbed on the metal, of the metal surface and of the free standing \MS, respectively. The system as a whole is neutral, and $\Delta{n(\mathbf{r})}$ is localized around the metal/\MS\ interface, i.e. $\Delta{n}(\mathbf{r})\rightarrow 0$ for $\mathbf{r}$ sufficiently far from the interface. Solving the Poisson equation with $\Delta{n}(\mathbf{r})$ as source then gives a potential step across the interface
\begin{equation}
\Delta{V} = \frac{e^2}{\epsilon_{0}A}\iiint z \Delta{n}(\mathbf{r})\,dxdydz. 
\label{eq:sb2}
\end{equation}
Here $z$ is the direction normal to the interface, $A$ is the interface area, and $\Delta V$ is the difference between the asymptotic values of the potential left and right of the interface. Figure~\ref{fig:deltaV} also illustrates an alternative expression for the potential step 
\begin{equation}
 \Delta{V} = W_\mathrm{M}  - W_\mathrm{ M \mid MoS_{2}}, 
\label{eq:sb3}
\end{equation}
where $W_\mathrm{M}$,  $W_\mathrm{ M \mid MoS_{2}}$ are the work functions of the clean metal surface, and of the metal surface covered by \MS, respectively. A practical way of obtaining work functions from DFT calculations is to track the plane-averaged electrostatic (Hartree) potential $\overline{V}(z)$ into the vacuum, see Fig.~\ref{fig:deltaV}, where typically the asymptotic value is reached with a few \AA\ from the surface. In converged calculations the expressions of Eqs. \ref{eq:sb2} and \ref{eq:sb3} give results that are with a few meV of one another.

\begin{figure*}[!tpb]
\begin{center}
\includegraphics[width=2.0\columnwidth]{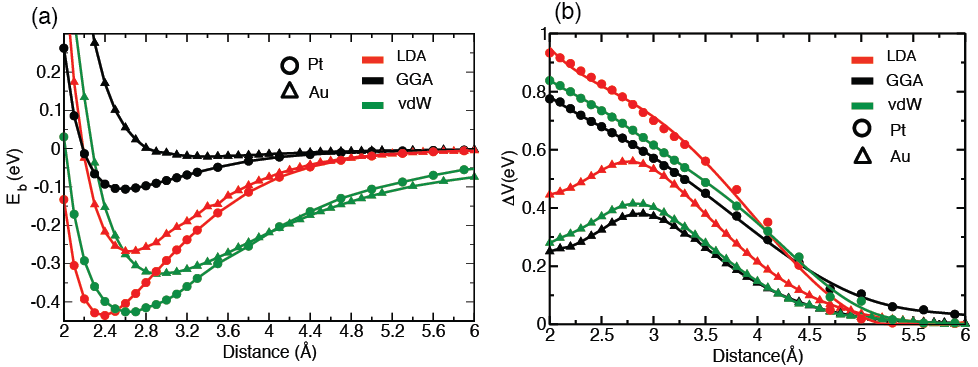}
\caption{(Color online) (a) Binding energy curves $E_b(d)$ of \MS\ on Au and Pt(111), calculated with the GGA/PBE (black), opt88b-vdw-DF (green), and LDA (red) functionals. (b) Interface potential steps $\Delta V(d)$ for \MS\ on Au and Pt(111), calculated with the three functionals. $d$ is the distance between the top metal and the bottom sulfur layers.}
 \label{fig:binding}
\end{center}
\end{figure*}

The interface potential step plays an important role in determining the Schottky barrier, i.e., the potential barrier for charge carrier transport from a metal into a semiconductor. The Schottky barrier for electrons can be written as
\begin{equation}
\Phi_\mathrm{n} = E_F - \chi_\mathrm{MoS_{2}} = W_\mathrm{M}  - \chi_\mathrm{MoS_{2}} - \Delta{V}, 
\label{eq:sb}
\end{equation}
with $E_F$ the Fermi level and $\chi_\mathrm{MoS_{2}}$ the electron affinity of \MS. There are several ways to extract the Schottky barrier height from \MS/metal slab calculations. 

One could determine $\Phi_\mathrm{n}$ by measuring $E_F - \chi_\mathrm{MoS_{2}}$ in the band structure or in the density of states of the \MS/metal slab, as in Refs.~\onlinecite{Chen:nanol13,Kang:prx14,Gong:nanol14} and \onlinecite{Li:acs23}. In order to identify the \MS\ related states, one needs to calculate the amplitudes of the projections of the wave functions of the slab on the \MS\ layer. There is always some arbitrariness involved in such a projection if the adsorbate and the substrate are in close connection. In addition, identification of states belonging to the adsorbate is possible only if its electronic structure is not significantly perturbed in the adsorption process, which is only the case if the adsorbate is (weakly) physisorbed on the substrate.\cite{Bokdam:prb14b} In practice we find that this procedure for obtaining the Schottky barrier height at \MS/metal interfaces is not sufficiently accurate when applied to the (projected) density of states, and of practical use only when applied to the (projected) band structure of a small supercell.

In contrast, starting from the right-hand side of Eq.~\ref{eq:sb}, $W_\mathrm{M}$,  $\chi_\mathrm{MoS_{2}}$ and $\Delta{V}$ are easily obtained in separate calculations on the clean metal surface, the free-standing \MS\ layer, and the \MS/metal slab, respectively. Of course, if the \MS\ electronic structure is very strongly perturbed by adsorption, one has to reconsider the definition of the Schottky barrier, see Sec.~\ref{sec:Ti}. In the following the potential step $\Delta V$ is used to characterize the \MS/metal interface, along with the binding energy and the structure.

An alternative way of locating the conduction band edge $\chi_\mathrm{MoS_{2}}$ in a \MS/metal slab calculation, without having to resort to wave function projections, is by aligning the core levels of the Mo or S atoms in the slab with the corresponding core levels in free-standing \MS. It allows us to compare the densities of states of free-standing and adsorbed \MS, see Sec.~\ref{sec:sb}. In principle, this procedure also allows for calculating the Schottky barrier height. As the method described in the previous paragraph, this only makes sense if the \MS\ electronic structure is not perturbed too strongly by the adsorption.

\subsection{Comparison of DFT Functionals}
\label{sec:functionals}

Materials such as graphite, \BN\,, and \MS\,have a layered structure, where the atoms within one layer form strong covalent bonds, but the interaction between the layers consists of weak, van der Waals, forces. Common GGA functionals, such as PBE,\cite{Perdew:prl96} lack a description of van der Waals interactions, which results in a severe underestimation of the interlayer binding energy in graphite and \BN, and an overestimation of the interlayer bonding distance.\cite{Sachs:prb84,Hazrati:prb14} Similar problems are encountered when graphene or \BN\,are adsorbed on a metal substrate.\cite{Laskowski:prb08,Stradi:prl11,Olsen:prl11,Andersen:prb12,Janthon:jcp13} The LDA functional also lacks a description of van der Waals interactions, but it, somewhat fortuitously, gives reasonable binding energies and geometries for graphite, \BN, and for the adsorption of these materials on metals.\cite{Giovannetti:prl08,Khomyakov:prb11,Bokdam:prb14a,Bokdam:prb14b} In general however, the LDA functional tends to overestimate binding energies, which is regularly accompanied by an underestimation of the bonding distance. 

\label{sec:functionals}
\begin{table*}
\caption{Equilibrium bonding distance $d_{eq}$, binding energy $E_{b}$, and interface potential step $\Delta V$, for \MS\ on metal (111) surfaces, calculated with different functionals.}
\begin{ruledtabular}
\begin{centering}
  \begin{tabular}{ccccccccccccc}

    \multirow{2}{*}{} &
      \multicolumn{3}{c}{Au} &
      \multicolumn{3}{c}{Ag} &
      \multicolumn{3}{c}{Pd} &
      \multicolumn{3}{c}{Pt} \\
      \cline{2-4}
      \cline{5-7}
      \cline{8-10}
      \cline{11-13}\\
    & $d_{eq}$(\AA) & $\Delta{V}$(eV) & $E_{b}$(eV) & $d_{eq}$(\AA) & $\Delta{V}$(eV)& $E_{b}$(eV) &  $d_{eq}$(\AA) & $\Delta{V}$(eV) & $E_{b}$(eV)&  $d_{eq}$(\AA) & $\Delta{V}$(eV)& $E_{b}$(eV)\\
    \hline
    LDA                         & 2.6 & 0.54   &$-$0.27      &2.5  & 0.10    &$-$0.33     &   2.2  &  0.50 & $-$0.69& 2.4 & 0.85 & $-$0.43     \\
    PBE              & 3.3 & 0.38 &$-$0.02     &2.8    & 0.10     & $-$0.08    &    2.3 &  0.34 & $-$0.25& 2.6 & 0.66 & $-$0.11      \\
    vdW-DF & 2.9 & 0.41 & $-$0.33    &2.8    & 0.11         & $-$0.35    &  2.4   &  0.30      &$-$0.54 & 2.6 & 0.71 & $-$0.43     \\
  \end{tabular}
  \par\end{centering}
\end{ruledtabular} 
\label{tab:functionals}
\end{table*}

Many of these problems are mitigated when using vdW-DFs,\cite{Bjorkman:prl108,Hamada:prb89} which, for instance, describe the bonding in graphite very well.\cite{Hazrati:prb14} 
The exchange-correlation energy in a vdW-DF takes the form
\begin{equation}
E_\mathrm{xc} = E_\mathrm{x}+E_\mathrm{c} ^\mathrm{vdW}+ E_\mathrm{c}^\mathrm{loc}, 
\label{eq:sb4}
\end{equation}
where $E_\mathrm{x}$, $E_\mathrm{c}^\mathrm{loc}$ and $E_\mathrm{c} ^\mathrm{vdW}$ describe the exchange part, and the local and nonlocal electron-electron correlations, respectively. For $E_\mathrm{c} ^\mathrm{vdW}$ we use the vdW kernel developed by Dion \textit{et al.} \cite{Dion:prl04} and for $E_\mathrm{c}^\mathrm{loc}$ the correlation part of the LDA functional. For the exchange part $E_\mathrm{x}$ we use the optB88 functional.\cite{Klimes:prb11} The opt88-vdW-DF has given good results for binding energies and geometries of graphite, \BN, and the adsorption of these materials on metals.\cite{Hazrati:prb14,Bokdam:prb14a}  

In the following we test the GGA/PBE, LDA, and opt88-vdW-DF functionals for the adsorption of \MS\ on metals. As test cases we use the 4d and 5d metals Ag, Au, Pd, and Pt. We place a \MS\ monolayer on top of the (111) surface of these metals, choosing a $\sqrt{3} \times \sqrt{3}R30^\mathrm{o}$ in-plane \MS\ unit cell on top of a $2\times 2$ (111) surface cell. The in-plane \MS\ lattice parameters are kept at their optimized values for a free-standing layer, and the in-plane metal lattice parameter is adapted accordingly. The size of the adaption is maximal for Au, where it results in a compression of the in-plane Au lattice by 4.2\%. The effects of this artificial strain are discussed in the next section.

Figure~\ref{fig:binding}(a) shows the binding curves of \MS\,on Au and Pt(111) for the three functionals. The binding energy is defined as the total energy per \MS\,formula unit of the metal/\MS\,slab minus the total energies of the clean metal slab and the free-standing \MS\,layer, as a function of the distance $d$ between the top layer of metal atoms and the bottom layer of sulfur atoms. For \MS\ on Au(111), PBE gives virtually no bonding, and opt88-vdW-DF gives a sizable binding energy. The opt88-vdW-DF result suggests that \MS\ is physisorbed on Au(111), with van der Waals interactions playing the decisive role in the bonding. PBE does not capture this at all. LDA gives a equilibrium binding distance that is 0.3 \AA\ smaller, and an equilibrium binding energy that is 32\% larger.

For \MS\ on Pt(111) all three functionals give equilibrium bonding distances that are shorter than for \MS\ on Au(111), and a bonding that is stronger, which suggests that \MS\ may be weakly chemisorbed on Pt(111). PBE and opt88-vdW-DF give a similar equilibrium distance, although PBE captures only 26\% of the binding energy, indicating that van der Waals interactions still play an important role here. LDA gives a similar binding energy as opt88-vdW-DF, but an equilibrium binding distance that is 0.2 \AA\ smaller.   

Table \ref{tab:functionals} shows the equilibrium binding distances and energies obtained with the three functionals for \MS\,on Au, Ag, Pd, and Pt(111). Treating the results for opt88-vdW-DF as a benchmark, PBE is seen to severely underestimate binding energies, whereas LDA gives quite reasonable binding energies. LDA however gives binding distances that are up to 0.3 \AA\,shorter than those obtained with opt88-vdW-DF, in particular for cases where the bonding is weak, such as Au and Ag. In contrast, PBE gives binding distances that are similar to those obtained with opt88-vdW-DF, except for Au, where PBE essentially fails to give any significant bonding.  

Potential steps $\Delta{V}$ as a function of the distance $d$ between the top layer of metal atoms and the bottom layer of sulfur atoms, calculated according to Eq.~\ref{eq:sb3}, are shown in Fig.~\ref{fig:binding}(b) for Au and Pt. The curves for the PBE and the opt88-vdW-DF functionals are within 0.05 eV of one another in the range $d=2.5$-3 \AA, whereas LDA gives a potential step that is 0.10-0.15 eV higher. In view of the considerable differences in the binding curves for these three functionals, the differences in the potential steps are remarkably small. This is true for all metal substrates listed in Table~\ref{tab:functionals}.

In Ref.~\onlinecite{Bokdam:prb14b} the main contribution to the potential step in the adsorption of \BN\ on metal substrates was attributed to Pauli repulsion. This can be modeled by an electron density that is obtained by anti-symmetrizing the product of the metal and the adsorbate wave functions. As long as these wave functions do not strongly depend on the functional, the electron density and the potential step are also relatively insensitive to the functional used. This is unlike the total energy, which for a given electron density is very dependent on the functional. For the potential step to be accurate it is however important to obtain the proper equilibrium binding distance.\cite{Giovannetti:prl08,Khomyakov:prb11,Bokdam:prb14a,Bokdam:prb14b} 


\subsection{Lattice Mismatch}
\label{sec:mismatch}
The absolute values of the binding energies given in Table~\ref{tab:functionals} are much smaller than what one expects to find for true chemical bonding. The differences between the values obtained with PBE and opt88-vdW-DF indicate that van der Waals interactions play a significant role in the bonding. With such a weak metal/adsorbate bonding it is unlikely that the metal substrate can enforce its lattice periodicity onto the \MS\ overlayer. Therefore, a metal/\MS\ interface very likely becomes incommensurable if the metal/\MS\ lattice mismatch is substantial. In electronic structure calculations one is forced to use commensurable structures to model incommensurable systems. Obviously care must be taken to ensure that the artificial strain introduced this way, does not alter the electronic structure in an unrealistic way. 

Based upon previous experience, we expect that modifying the in-plane lattice constant of a close-packed metal surface by a few percent only affects its electronic properties mildly.\cite{Giovannetti:prl08,Khomyakov:prb11,Bokdam:prb14a,Bokdam:prb14b} In contrast, changing the lattice parameter of \MS\ by just one percent already alters the band gap by $\sim$0.1 eV, and changes it from direct to indirect. A larger change in the lattice parameter has an even more dramatic effect. Applying a tensile strain of $\sim 5$\% to \MS\ reduces the band gap by $\sim 1$ eV.\cite{Shi:prb13,LiTian:prb85,WonSeok:prb85,Cappel:prb88,Thygesen:prb88,Conley:nanol13}

\begin{figure}[tb]
\includegraphics[width=0.5\textwidth]{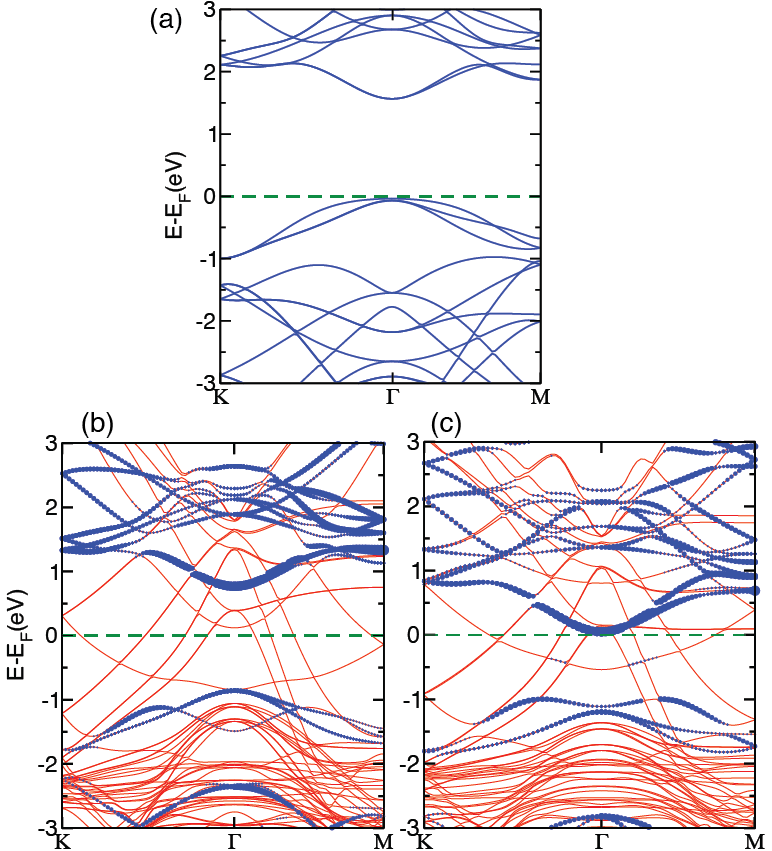}
\caption{(Color online) (a) Band structure of a free-standing \MS\ monolayer in a $\sqrt{3}$$\times$$\sqrt{3}$ cell, where the direct band gap appears at $\Gamma$; (b) Band structure of \MS/Au(111) with the in-plane Au lattice compressed by 4.2\% to match the \MS\ lattice; the blue color indicates the weight of a projection of the wave functions on the \MS\ sites; (c) as (b) but with the \MS\ lattice stretched by 4.2\% to match the Au(111) lattice.} 
\label{ref:strain}
\end{figure}

As an example, the PBE optimized in-plane lattice parameters of \MS\ and Au(111) are 3.19 \AA\,and 2.88 \AA. Placing a $(\sqrt{3}\times\sqrt{3})R30^{\circ}$ \MS\ cell on top of 2$\times$2 Au(111) surface cell then leads to a lattice mismatch of 4.2\%. Figure~\ref{ref:strain}(b) shows the electronic band structure of \MS/Au(111) where the in-plane Au(111) is compressed by 4.2\% to match the lattice parameter of \MS. As the interaction between \MS\ and the Au surface is relatively small, it is not surprising to see that the band structure of adsorbed \MS\ resembles that of free-standing \MS, shown in Fig.~\ref{ref:strain}(a). Note that in the $\sqrt{3}\times\sqrt{3}$ \MS\ cell the bands are folded such that the direct band gap appears at the $\Gamma$ point. The work function of clean Au(111) is changed by only 0.08 eV by the 4.2\% compression of its lattice.

For comparison, Fig.~\ref{ref:strain}(c) shows the band structure of \MS/Au(111) when \MS\ is stretched by 4.2\% to match the Au(111) lattice. Clearly the band structure of \MS\ is now changed significantly. It no longer shows a direct band gap at $\Gamma$, but an indirect band gap, and the size of the band gap is reduced to $\sim$1 eV, which is consistent with previous studies.\cite{Shi:prb13,LiTian:prb85,WonSeok:prb85,Cappel:prb88,Thygesen:prb88,Conley:nanol13} The Schottky barrier for electrons (the energy difference between the bottom of the conduction band and the Fermi level), which is a sizable 0.7 eV in Fig.~\ref{ref:strain}(b), is reduced to zero in Fig.~\ref{ref:strain}(c) as in Ref.~\onlinecite{Popov:prl108}. The latter is clearly unphysical: one would not expect a high work-function metal such as Au to form a barrierless contact for electrons. Indeed experimentally Au is found to form a substantial Schottky barrier with \MS.\cite{Lince:prb87,Liu:acsnano12,Fontana:scirep13,Kaushik:apl14,Maurel:ss05}

In the following we base the in-plane lattice constant of the \MS/metal slab on the optimized values of free-standing \MS,  which are 3.13, 3.18, and 3.19 \AA\,for the LDA, optb88-vdW-DF, and PBE functionals, respectively. Experimentally reported bulk \MS\ lattice constants are in the range 3.13-3.16 \AA,\cite{Young:jpd68,Alhili:jcg15,Boker:prb64} suggesting that the LDA result may be more accurate and both PBE and the vdW functional are overestimating the lattice constant somewhat.

In making a commensurable structure we adapt the metal to the \MS\ lattice. To minimize the artificial strain that is introduced by this adaptation, we construct in-plane supercells following the procedure of Ref.~\onlinecite{Komsa:prb88}. We denote a basis vector of a \MS\ supercell by $\vec{T}_{1}=n_{1}\vec{a}_{1}+n_{2}\vec{a}_{2}$, with $\{\vec{a}_{1},\vec{a}_{2}\}$ the basis vectors of the primitive cell, and $n_1,n_2$ integers. Similarly, $\vec{T}'_{1}=m_{1}\vec{b}_{1}+m_{2}\vec{b}_{2}$ is a basis vector of a metal surface supercell, with $\{\vec{b}_{1},\vec{b}_{2}\}$ the basis vectors of the primitive cell, and $m_1,m_2$ integers. We search for a set of values for $n_{1}$,$n_{2}$,$m_{1}$, and $m_{2}$, such that the difference between the \MS\ and the metal supercell basis vectors is less than a margin $\delta$,
\begin{equation}
 \frac{\vert \vec{T}_{1} \vert - \vert \vec{T}'_{1} \vert}{\vert \vec{T}_{1} \vert} \leq \delta.
\label{eq:sb6}
\end{equation}
We then rotate the \MS\ lattice by an angle $\alpha$ such, that the directions of the $\vec{T}_{1}$ and $\vec{T}'_{1}$ vectors coincide. Because of the symmetry of the lattice the second basis vector of the supercell is easily obtained by a 120$^\mathrm{o}$ rotation, $\vec{T}_{2}=R(120^\mathrm{o})\vec{T}_{1}$. The commonly used surface science notation of this supercell is a $\sqrt{N}\times \sqrt{N}R\alpha$ \MS\ lattice on top of a $\sqrt{M}\times \sqrt{M}$ metal lattice, where $N=n_1^2+n_2^2+n_1n_2$ and $M=m_1^2+m_2^2+m_1m_2$.

\begin{table}[tb]
\caption{ In-plane supercell defined by the \MS\ lattice vector $R(\alpha)\vec{T}_{1}$, where  $\vec{T}_{1}=n_{1}\vec{a}_{1}+n_{2}\vec{a}_{2}$  and the metal lattice vector $\vec{T}'_{1}=m_{1}\vec{b}_{1}+m_{2}\vec{b}_{2}$. $\delta$ gives the mismatch between the \MS\ and metal lattices, Eq.~\ref{eq:sb6} (PBE values).}
\begin{ruledtabular}
\begin{centering}
\begin{tabular}{lcccc}
         &  $n_{1},n_{2}$ & $m_{1},m_{2}$ & $\alpha$ & $\delta$ (\%) \\
\hline
Mg           &  $1,0$ & $1,0$ & $0^\mathrm{o}$                       & 0.6       \\
Al             &  $4,-1$ & $4,0$ & $13.9^\mathrm{o}$      & 0.5       \\
Ag           &  $4,-1$ & $4,0$ & $13.9^\mathrm{o}$      & 0.15     \\
Ti           &  $5,-2$ & $4,0$ & $23.4^\mathrm{o}$      & 0.7     \\
Cu          &  $4,0$ & $5,0$ & $0^\mathrm{o}$             & 0.3       \\
Au           &  $4,-1$ & $4,0$ & $13.9^\mathrm{o}$      & 0.15     \\
Pd          &  $1,1$ & $2,0$ & $30^\mathrm{o}$           & 0.3        \\
Pt           &  $1,1$ & $2,0$ & $30^\mathrm{o}$           & 0.3        \\
Co		&  $5,-4$ & $4,-3$ & $3^\mathrm{o}$           & 0.01        \\
Ni		&  $5,-4$ & $4,-3$ & $3^\mathrm{o}$           & 0.8        
\end{tabular}
\par\end{centering}
\end{ruledtabular} 
\label{tab:supercell}
\end{table}

\begin{figure}[tb]
\includegraphics[width=0.8\columnwidth]{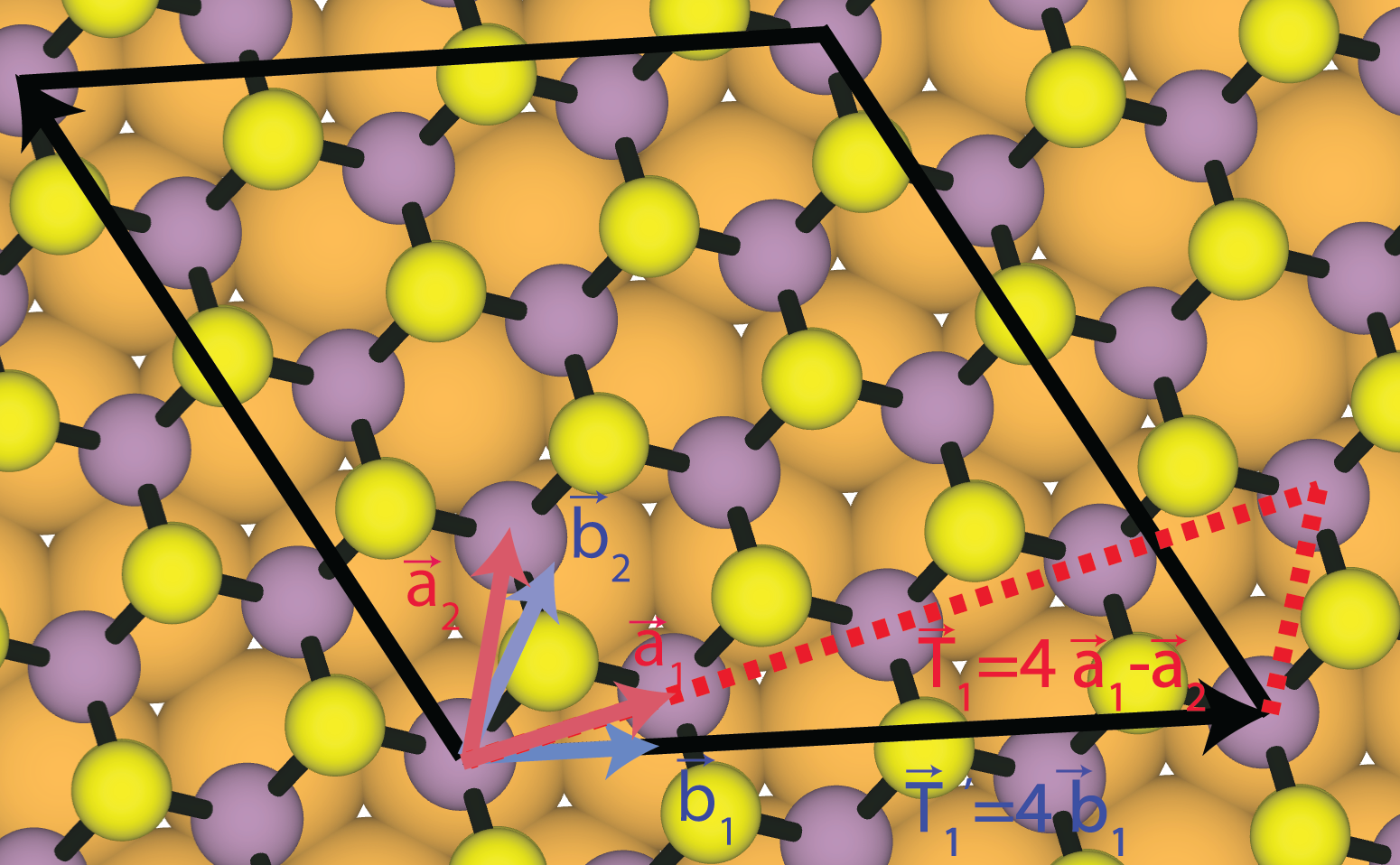}
\caption{(Color online) Top view of \MS/Au(111) interface indicating the supercell (black lines), the primitive basis vectors $\vec{a}_{1},\vec{a}_{2}$ and $\vec{b}_{1},\vec{b}_{2}$ of the \MS\ and Au(111) lattices, respectively, and the basis vector $\vec{T}_{1}$ and $\vec{T}'_{1}$ of the supercell.}
 \label{fig:Au_supercell}
\end{figure}

The parameter $\delta$ determines the mismatch between the \MS\ and the metal lattices, and the strain we apply to the metal lattice. In this study, we choose the smallest supercell for which $\delta < 1\%$. Figure~\ref{fig:Au_supercell} gives an example of a supercell for \MS\ on Au(111) that is constructed this way, and Table~\ref{tab:supercell} lists the supercells and the lattice mismatch $\delta$ used in this study for the different metals.

\begin{figure}[tb]
\includegraphics[width=0.9\columnwidth]{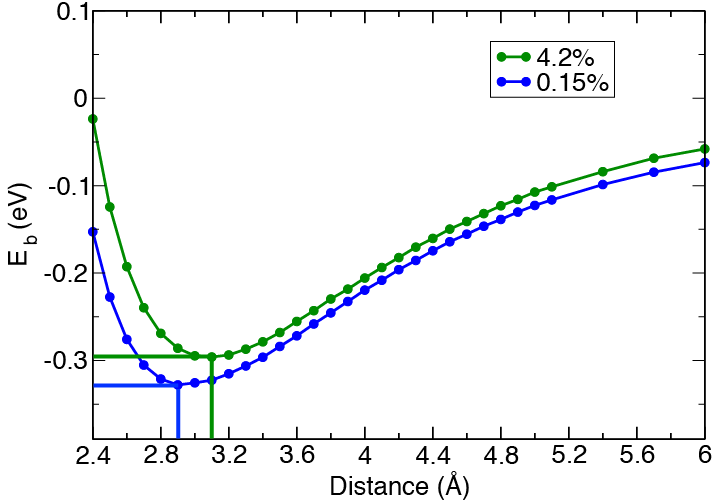}
\caption{(Color online) Binding energy curves $E_b(d)$(eV) of \MS\ on Au(111), calculated with opt8b-vdW-DF functional for a mismatch of 0.15$\%$ (blue), and 4.2$\%$ (green) between the \MS\ and the Au(111) lattices.} 
\label{fig:binding_mismatch}
\end{figure}

\begin{table}[tb]
\caption{Equilibrium bonding distance $d_{eq}$, binding energy $E_{b}$, and interface potential step $\Delta V$, for \MS\ on metal (111) surfaces, calculated with supercell lattices with a different mismatch $\delta$.}
\begin{ruledtabular}
  \begin{tabular}{ccccccccccccc}

    \multirow{2}{*}{$\delta(\%)$ } &
      \multicolumn{3}{c}{Au} &
      \multicolumn{3}{c}{Ag}  \\
      \cline{2-4}
      \cline{5-7}\\
    & $d_{eq}(\AA)$ & $\Delta{V}(eV)$ & $E_{b}(eV)$ & $d_{eq}(\AA)$ & $\Delta{V}(eV)$& $E_{b}$ \\
    \hline
    $0.15$                         & 2.9   & 0.41   &$-0.33$   &2.8   & 0.11    & $-0.35$    \\
    $4.2$                           & 3.1 & 0.51 & $-0.30$   &2.9     & 0.47 & $-0.32$       \\
  \end{tabular}
\end{ruledtabular} 
\label{tab:compress}
\end{table}

In the calculations discussed in Secs.~\ref{sec:functionals} and \ref{sec:mismatch} we have used a $\sqrt{3} \times \sqrt{3}R30^\mathrm{o}$ \MS\ cell on top of a $2\times 2$ Au(111) cell, which leads to a lattice mismatch of 4.2\%. A $\sqrt{13} \times \sqrt{13}R13.9^\mathrm{o}$ on top of a $4\times 4$ Au supercell, see Table~\ref{tab:supercell} and Fig.~\ref{fig:Au_supercell}, reduces the lattice mismatch to 0.15\%. Figure~\ref{fig:binding_mismatch} shows that the binding energy curves for the two structures are quite similar. The equilibrium binding energy is increased by 0.03 eV upon compressing the Au lattice by 4.2\%, and the equilibrium binding distance is decreased by 0.02 \AA. Typically the interface potential step is affected by the compression on a scale of 0.1 eV, as is shown in Table~\ref{tab:compress}. However sometimes the effect is larger, as for Ag. In conclusion, compressing the metal lattice does not generally have the same dramatic effect as stretching the \MS\ lattice has, but large lattice mismatches should be avoided.


%
%
%
%

\section{RESULTS}
\label{sec:results}
\subsection{Metal/\MS\ interaction}
\label{sec:bonding}

Calculated equilibrium binding energies and bonding distances for the \MS/metal structures of Table~\ref{tab:supercell} are listed in Table~\ref{tab:energies}. The binding energies obtained with opt88-vdW-DF are in the range $-0.3$ to $-0.6$ eV. These numbers seem somewhat too low in order to classify the bonding as physisorption, yet too high to call it chemisorption. Van der Waals interactions play an important role in the bonding, which becomes especially clear when comparing to the results obtained by PBE. The PBE functional lacks van der Waals interactions, and it typically captures only approximately half the \MS/metal binding energy or less.      

\begin{table}
\caption{Equilibrium  binding energy $E_{b}$, and bonding distance $d_{eq}$, for \MS\ on metal (111) and (0001) surfaces in the interface structures of Table~\ref{tab:supercell}, calculated with the optb88b-vdW-DF and the PBE functionals.}
\begin{ruledtabular}
\begin{centering}
\begin{tabular}{lccccccccc}
          &$E_\mathrm{b(vdW)}$&$d_\mathrm{{eq}(vdw)}$ & $E_\mathrm{b(PBE)}$ & $d_\mathrm{{eq}(PBE)}$  \\
                                 &      (eV)                 & (\AA)          & (eV) & (\AA) \\
\hline
Mg           & $-0.55$          &2.3   &  $-0.20$      & 2.2               \\
Al            & $-0.30$           &2.8   & $-0.30$       &2.8                \\
Ag           & $-0.35$         & 2.8   &  $-0.08$        &2.9                \\
Ti            & $-0.51$          &2.3   &  $-0.67$      & 2.3               \\
Cu          &  $-0.40 $        &2.5    &  $-0.16$      & 2.4                    \\
Au          &  $-0.33$         &2.9   &  $-0.02$       & 3.3               \\
Pd          &  $-0.54$         &  2.4  &  $-0.25$      & 2.3              \\
Ni		& $-0.51$          & 2.2  &  $-0.25$     & 2.2              \\
Co		& $-0.57$          & 2.2  &  $-0.29$       & 2.2               \\
Pt           & $-0.43$          & 2.6   &  $-0.11$      & 2.6               \\
\end{tabular}
\par\end{centering}
\end{ruledtabular} 
\label{tab:energies}
\end{table}

A noticeable exception is \MS/Ti(0001), where PBE gives approximately double the opt88-vdW-DF binding energy. It suggests that \MS\ is chemisorbed on Ti(0001), which is described better by PBE. This case will be discussed in more detail in Sec.~\ref{sec:Ti}. In contrast, the PBE functional essentially fails to give bonding for the adsorption of \MS\ on Au(111), and all bonding comes from van der Waals interactions, so we may classify this case as physisorption. For the other metals it is difficult to make a distinction between physisorption and chemisorption on the basis of the binding energy alone. 

\begin{figure}
\includegraphics[width=1.0\columnwidth]{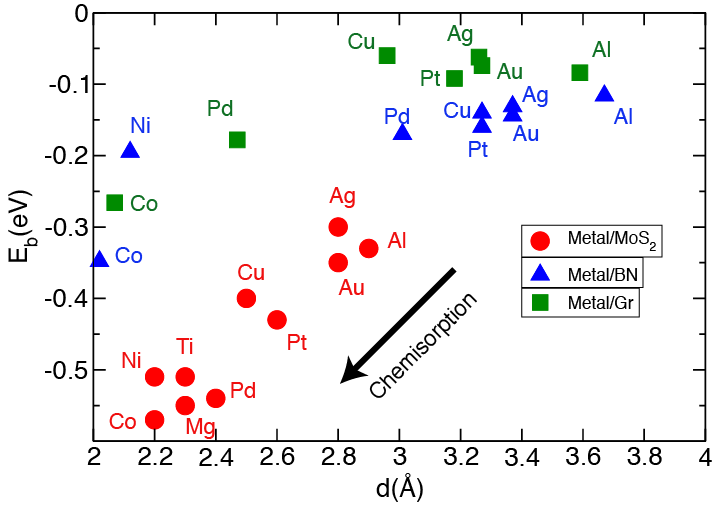}
\caption{(Color online) Binding energy $E_{b}$(eV) per \MS\ versus equilibrium bonding distance $d_\mathrm{eq}$(\AA) for \MS\ adsorbed on metal(111) and (0001) substrates (red circles), as calculated with the optb88-vdW-DF functional. For comparison, results for \BN\ (blue triangles),\cite{Bokdam:prb14a} and graphene (green squares),\cite{Gong:JAP} are also shown.}
 \label{fig:physchem}
\end{figure}

In general terms, physisorption is accompanied by a weak perturbation of the electronic structure of the adsorbed layer, whereas chemisorption results in a sizable perturbation of that electronic structure. For graphene and \BN\ adsorbed on metal surfaces it was possible to correlate that perturbation with the equilibrium bonding distances $d_\mathrm{eq}$. Those distances can be divided into two groups separated by a critical binding distance $d_\mathrm{c}$. For $d_\mathrm{eq}>d_\mathrm{c}$, the bonding is physisorption, and for $d_\mathrm{eq}<d_\mathrm{c}$, the bonding is chemisorption. For graphene and \BN\ this distinction is successful because there are hardly any cases where $d_\mathrm{eq}\approx d_\mathrm{c}\approx 2.8$ \AA\, as is illustrated in Fig.~\ref{fig:physchem}. Clearly bonding distances and energies are correlated; a shorter distance generally gives a lower energy.

Plotting the binding energies and distances for \MS/metal interfaces in Fig.~\ref{fig:physchem}, one observes that the distinction between physisorption and chemisorption is much less clear for this case. The binding of \MS\ to a metal substrate is stronger than that of graphene or \BN, reflecting the fact that van der Waals interactions increase with the atomic number. Maybe somewhat surprisingly the bonding distance of \MS\ to a metal substrate is generally shorter than that of graphene or \BN. Graphene and \BN\ have $\pi$-orbitals that stick out below their respective planes, which give rise to a substantial Pauli repulsion at distances to the metal plane of $\lesssim 3$ \AA.\cite{Bokdam:prb14b}  Apparently the wave functions of \MS\ do not stick out that far below the plane of the bottom sulfur layer.  

\begin{figure}
\includegraphics[width=1.0\columnwidth]{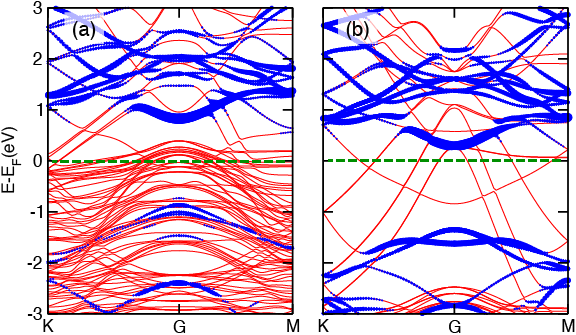}
\caption{(Color online) (a) Band structure of \MS/Pd(111); the blue color indicates the weight of a projection of the wave functions on the \MS\ sites; (b) idem for \MS/Ag(111); for comparison both band structures are shown in a $\sqrt{3}$$\times$$\sqrt{3}$ surface cell.}
 \label{fig:AgPd}
\end{figure}

The bonding distances for \MS/metal interfaces cannot easily be simply into two groups, as is the case for graphene and \BN/metal interfaces. Instead there is a more gradual scale. The bonding distances of \MS\ on Al, Au and Ag are on the physisorption side of Fig.~\ref{fig:physchem}, whereas on Co, Ni, Mg, and Ti, they are more on the chemisorption side, with Pt, Cu, Pd as intermediate cases. However, a clear dividing line like for graphene and \BN\ can not be drawn. Indeed if one considers the \MS/metal interface for two similar metals that give rise to a fairly large difference in bonding distance and binding energy: Ag and Pd, one does not observe a qualitative difference in the the electronic structure of the \MS\ adsorbate, see Fig.~\ref{fig:AgPd}. In both cases the \MS\,bands are perturbed by the metal-\MS\ interaction, but the signature of the \MS\ bands can still be recognized. In particular, it still seems to be possible to identify the top of the \MS\ valence band, and the bottom of the conduction band. Nevertheless the \MS\ states do hybridize with those of the metal substrate, as we will discuss in the next section.

Another way of characterizing the bonding is to analyze the geometry of the \MS/metal interface. Chemisorption involves the formation of chemical bonds between the adsorbate and the metal, which frequently also leads to a deformation of the adsorbate's structure. We can define a displacement $\Delta_i = \vert \mathbf{R_i} - \mathbf{R}_{0,i} \vert$ between the position $\mathbf{R_i}$ of an atom $i$ in the optimized \MS/metal structure, and its position $\mathbf{R}_{0,i}$ in the free-standing \MS\ or in the clean metal substrate. The displacements are obtained in a two-step procedure. First the \MS/metal structure is optimized while freezing the \MS\ layer and the metal substrate in their free-standing geometries. Once the equilibrium distance $d_\mathrm{eq}$ is obtained, as in Figs.~\ref{fig:binding} and \ref{fig:binding_mismatch}, all atomic positions are relaxed, and this last step defines the displacements. Table~\ref{tab:energies} gives $\overline{\Delta}_\mathrm{S}$ and $\overline{\Delta}_\mathrm{M}$, which are the average displacements of the bottom layer of sulfur atoms, and of the top layer of metal atoms, respectively, for some representative metal substrates. In addition, this table gives the maximum displacements $\Delta^\mathrm{max}_\mathrm{S}$ and $\Delta^\mathrm{max}_\mathrm{M}$.

\begin{table}
\caption{The average displacements $\overline{\Delta}_\mathrm{S}$ and $\overline{\Delta}_\mathrm{M}$ of the bottom layer of sulfur atoms, and of the top layer of metal atoms, and the corresponding maximum displacements $\Delta^\mathrm{max}_\mathrm{S}$ and $\Delta^\mathrm{max}_\mathrm{M}$, after the adsorption.}
\begin{ruledtabular}
\begin{centering}
\begin{tabular}{lccccccccc}
          & $\overline{\Delta}_\mathrm{S}$(\AA) &$\Delta^\mathrm{max}_\mathrm{S}$(\AA) &$\overline{\Delta}_\mathrm{M}$(\AA)& $\Delta^\mathrm{max}_\mathrm{M}$(\AA)\\
\hline
Ti            &  0.101  	 &0.152		& 0.176		 & 0.483             \\
Co		&0.010		&0.021		&0.068		&0.102\\
Pt 		&0.006		&0.009		&0.049		&0.098 \\
Ag 		&0.002		&0.005   	&0.030		&0.058\\
Au          & 0.001      	&0.004		&0.019           & 0.044   \\
Pd		&0.002		&0.003		&0.026		&0.051\\
Ni		&0.017		&0.034		&0.089		&0.150\\
Al		&0.001		&0.047		&0.046		&0.105\\
Mg		&0.022		&0.022		&0.091		&0.091\\
Cu		&0.033		&0.052		&0.070		&0.124\\
\end{tabular}
\par\end{centering}
\end{ruledtabular} 
\label{tab:structures}
\end{table}

The displacements are quite large for the \MS/Ti(0001) structure, indicating that there is a significant distortion of the geometries of both the \MS\ adsorbate and the Ti surface, which strongly suggests that \MS\ is chemisorbed on Ti. At the opposite end of the scale we find \MS/Au(111), where the atomic displacements are small, indicating that here we are in the physisorption regime. The behavior of the other metal substrates is in between these two extreme cases but more to the physisorption side. The 3d transition metals Co,Ni,Cu and the low work function simple metal Mg show somewhat larger distortions than the 4d and 5d metals Pd,Ag,Pt and the simple metal Al. 

\subsection{Interface Potential Step and Schottky Barrier}
\label{sec:sb}

Table~\ref{tab:DeltaV} gives the interface potential steps $\Delta V$ created by the adsorption of \MS\ on a metal substrate. This potential step strongly influences the Schottky barrier at metal/\MS\ contacts, see Eq.~\ref{eq:sb}, and as such it plays an important role in the physics of \MS\ semiconductor devices. The potential steps can be divided into two groups, i.e., positive $\Delta V$ for metals with a high work function, and negative $\Delta V$ for low work function metals. A positive $\Delta V$ means that  adsorption of \MS\ effectively lowers the work function of the substrate. The \MS\ layer has no intrinsic dipole moment perpendicular to the layer that could create such a potential step. So the work function lowering is a purely electronic effect that results from the displacement of surface electron density into the metal by physisorption of the adsorbate. 

\begin{table}[tb]
\caption{ Metal work function $W_\mathrm{M}$, interface potential step $\Delta V$, and Schottky barrier height $\Phi_\mathrm{n}$ calculated with  the PBE and opt88-vdW-DF functionals, with calculated \MS\ electron affinities of $\chi = 4.30$ eV and $\chi = 4.57$ eV, respectively.}
\begin{ruledtabular}
\begin{centering}
\begin{tabular}{lccccccc}
   & \multicolumn{2}{c}{PBE} & \multicolumn{4}{c}{vdW-DF} \\
   \cline{2-4} \cline{5-7} \\
         &$W_\mathrm{M}(eV)$ & $\Delta V(eV)$ & $\Phi_\mathrm{n}(eV)$ & $W_{M}(eV)$ & $\Delta V(eV)$ & $\Phi_\mathrm{n}(eV)$ \\      
\hline
Mg    & $3.78$       &   $-0.77$  & $0.25$     	& 3.96 & $-0.74$ & $0.13$   \\
Al      & $4.00$      &   $-0.54$   &$0.24$   	& 4.20 & $-0.56$  &  $0.19$  \\
Ag     & $4.47$      &  $0.10$     &$0.07$        & 4.82 &  $0.11$   &  $0.14$  \\
Ti      & $4.52$      &  $-0.28$     & 0.53$^\mathrm{a}$  	& 4.80 &  $-0.27$ &  0.54$^\mathrm{a}$  \\
Cu    & $4.70$      &  $0.35$      &$0.05$   	& 5.10 &   0.39     &   $0.14$ \\
Au    & $5.30$      &  $0.32$      &$0.68$  	& 5.58 &  $0.41$   &    $0.60$   \\
Pd    & $5.35$      &  $0.35$      &$0.70$  	& 5.48 &  $0.30$   &   $0.61$   \\
Pt     & $5.75$      &  $0.64$      &$0.81$ 		& 5.96 &  $0.68$   &   $0.71$\\
Co     & $5.13$      &  $0.29$     &$0.54$  	& 5.42 &  $0.34$   &   $0.51$\\
Ni     & $5.17$      &  $0.28$      &$0.59$		& 5.40 &  $0.37$   &   $0.46$\\
\end{tabular}
\par\end{centering}
\end{ruledtabular} 
$^\mathrm{a}$ See Sec.~\ref{sec:Ti}.
\label{tab:DeltaV}
\end{table}

This effect is known as the push-back effect or the pillow effect, which is a general phenomenon observed in the physisorption of closed-shell atoms, molecules, and layers on metal substrates. In Ref.~\onlinecite{Bokdam:prb14b} we have developed a quantitative model for this effect, based upon an anti-symmetrization of the product of the metal and adsorbate wave functions. When an adsorbate approaches a metal surface, the wave functions of the two systems overlap. Pauli exchange repulsion between these states leads to a spatial redistribution of the electron density, in particular to a decrease of the density in the overlap region. Since the metal wave functions are usually more extended and more easily deformable than those of the adsorbate, the net result of this redistribution is that electrons are pushed back into the metal, which effectively lowers the work function.

In the adsorption of graphene and \BN\,on high work function metals we found potentials steps of up to 1-2 eV. The potential steps for \MS\ adsorbed on the same metals are generally smaller, and more typically around 0.3-0.4 eV. The wave functions of first-row elements (B,C,N) are compact and not easily deformable, as compared to the wave functions of the metal substrate. The effect of Pauli repulsion in the metal/adsorbate overlap region is then very asymmetric. It is foremost the metal electron density that is deformed, i.e. pushed back, which gives a large work function lowering. If the adsorbate contains heavier elements, such as \MS, the effect of Pauli repulsion is more symmetric, i.e., both the metal and the adsorbate electrons are pushed out of the overlap region in a more symmetric way. This gives a smaller effect on the work function. Note that if the effect of Pauli repulsion would be completely symmetric, the work function would be unchanged. 

\begin{figure}[tb]
\includegraphics[width=1.0\columnwidth]{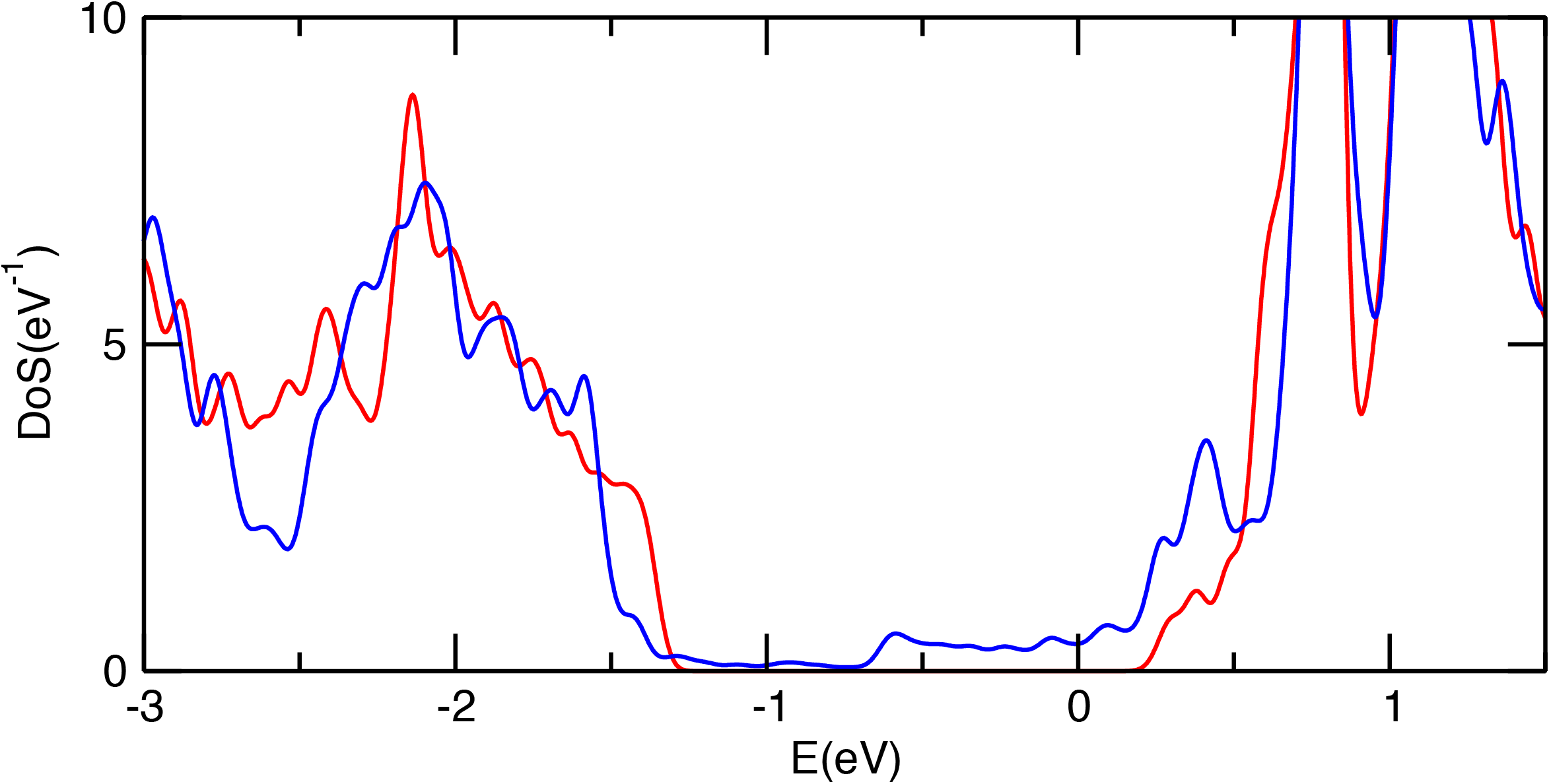}
\caption{(Color online) (blue) The total density of states of the \MS/Al(111) slab; (red) the sum of the densities of states of free-standing \MS\ and of the Al(111) slab. The densities of states are aligned by aligning the Mo 4s core levels, and the Al 2p core levels.}
 \label{fig:DOS_Al}
\end{figure}

Low work function metals experience an increase of the work function upon adsorption of \MS, i.e. a negative $\Delta V$, which indicates a net transfer of electrons from the metal to the \MS\ adsorbate. As \MS\ is a semiconductor it can only receive electrons in its conduction band. Therefore, for low work function metal substrates one expects the Fermi level to be in the conduction band of \MS. Analysis of the electronic structure of the \MS/metal slab however shows that this is not the case. The interaction between \MS\ and the metal at the interface leads to interface states with energies in the \MS\ band gap. That seems obvious if \MS\ is chemisorbed onto the substrate, as in the case of \MS/Ti(0001), which we will discuss in the next section.

\begin{figure}[!tpb]
\centering
\includegraphics[width=1.0\columnwidth]{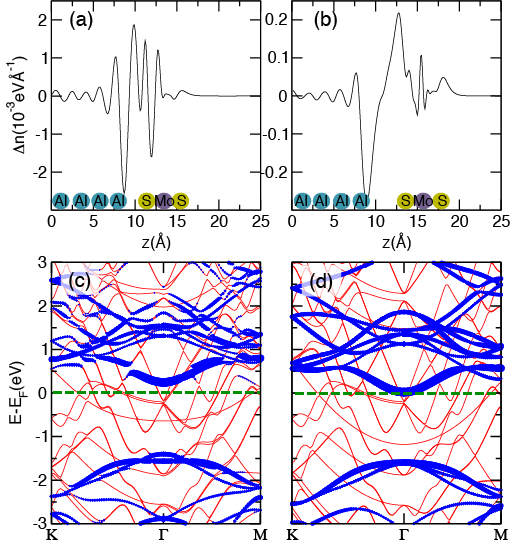}
\caption{(Color online) (a) The electron density difference $\Delta{n}(z)$ of the \MS/Al(111) interface at the equilibrium distance $d_{eq}=3.2$ \AA, and (b) at a distance $d=6$ \AA; (c,d) the corresponding band structures; the blue color indicates the weight of a projection of the wave functions on the \MS\ sites.}
 \label{fig:AlFinal}
\end{figure}

Somewhat surprisingly, a significant density of interface states also forms if the interaction between \MS\ and the metal substrate is relatively weak. For example, as discussed in the previous section, the interaction between \MS\ and Al(111) can be classified as physisorption, see Fig.~\ref{fig:physchem}. Nevertheless, states with energies inside the \MS\ band gap are formed at the \MS/Al(111) interface, as is immediately obvious when comparing the density of states of the interface with that of free-standing \MS, see Fig.~\ref{fig:DOS_Al}. The density of these interface states is not extremely high, yet sufficiently high to pin the Fermi level below the \MS\ conduction band, as demonstrated by Fig.~\ref{fig:DOS_Al}. The density of interface states increases with increasing \MS/metal interaction, but even for physisorption it seems sufficiently high to prevent the Fermi level from reaching the \MS\ conduction band. 

One can prove that these interface states are indeed responsible for pinning the Fermi level by artificially enlarging the distance between the \MS\ layer and the metal surface. This breaks the direct \MS/metal contact that is responsible for the formation of interface states. In the absence of interface states, the Fermi level is at the bottom of the \MS\ conduction band, see Fig.~\ref{fig:AlFinal}(d), which is what one would expect if the work function of the metal $W_\mathrm{M}$ is smaller than the electron affinity of \MS\ $\chi_\mathrm{MoS_2}$, see Eq.~\ref{eq:sb}. A transfer of electrons between the metal and the \MS\ overlayer then yields a charge distribution that can be associated with a simple interface dipole, see  Fig.~\ref{fig:AlFinal}(b). In contrast, if interface states are formed, the band structure of adsorbed \MS\ is perturbed, see  Fig.~\ref{fig:AlFinal}(c), and the pattern of the charge distribution at the interface is much more complicated, as shown in  Fig.~\ref{fig:AlFinal}(a). In that case, the interface states pin the Fermi level below the bottom of the \MS\ conduction band, see Fig.~\ref{fig:DOS_Al}. 

Schottky barrier heights (SBHs) for electrons, calculated according to Eq.~\ref{eq:sb}, are also listed in Table~\ref{tab:DeltaV}. The functional causes some uncertainty, as the work functions of the clean metal surfaces obtained with the opt88-vdW-DF functional tend to be somewhat higher than those obtained with the PBE functional. LDA in general gives even higher work functions, so opt88-vdW-DF gives work functions that are in between those of PBE and LDA.\cite{Khomyakov:prb11,Bokdam:prb14a} Note that the interface potential steps $\Delta V$ do not depend strongly on the functional. As the opt88-vdW-DF functional also gives a larger electron affinity for \MS, the Schottky barrier $\Phi_\mathrm{n}$ according to Eq.~\ref{eq:sb}, also does not depend strongly on the functional. 

We see that the SBH decreases with decreasing metal work function but does not go to zero. Instead it goes through a minimum for Cu and Ag, and then increases again for low work function metals like Al and Mg. As discussed above, this phenomenon is caused by interface states. The only way to get rid of such states is to break the direct interaction between \MS\ and the metal substrate. Ref.~\onlinecite{Farmanbar:prb15} discusses a practical way of doing this by inserting an atomic layer between the metal surface and the \MS\ layer. If this intermediate layer is purely van der Waals-bonded to \MS, no gap states are formed at its interface with \MS. In addition the intermediate layer should be transparent to electrons, such that the interface resistance is not dramatically increased. A monolayer of \BN\ or graphene satisfy these criteria.\cite{Farmanbar:prb15,WeiSun:acs9,Yuchen:IEE}

\subsection{\MS/Ti(0001)}
\label{sec:Ti}

\begin{figure}
\includegraphics[width=0.4\columnwidth]{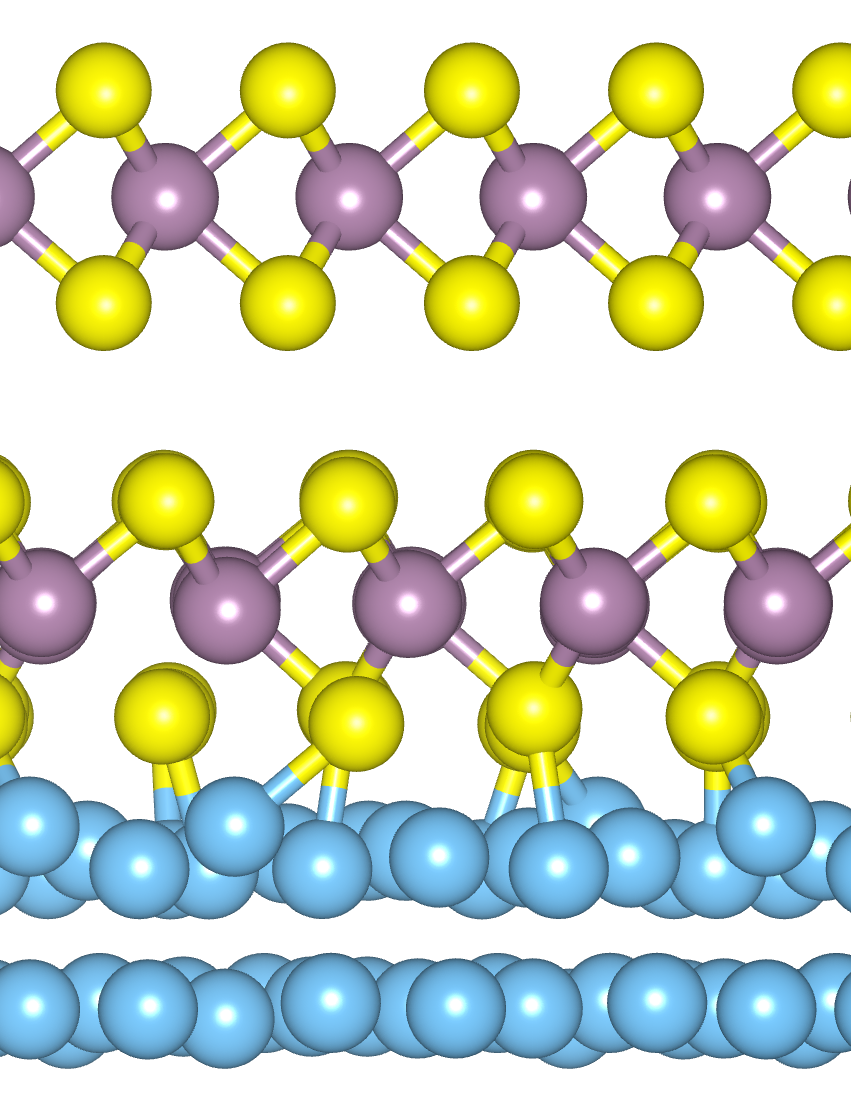}
\caption{(Color online) Side view of two layers of \MS\ adsorbed on Ti(0001).}
 \label{fig:Ti_structure}
\end{figure}

As discussed in Sec.~\ref{sec:bonding}, \MS\ is chemisorbed in Ti(0001). The binding energy and the equilibrium bonding distance of \MS\ on Ti(0001) does not seem to be qualitatively different from those for \MS\ on metal substrates such as Co(0001) or Ni(111), see Table~\ref{tab:energies} and Fig.~\ref{fig:physchem}. However, the structural deformation of the \MS\ layer adsorbed on Ti(0001) is much larger than that of \MS\ on other metals, see Table~\ref{tab:structures}. This deformation is illustrated in Fig.~\ref{fig:Ti_structure}. Atoms of the bottom sulfur layer make a bond with Ti atoms of the top layer of the substrate, where several of these metal atoms are pulled up from the substrate. The \MS\ and the Ti(0001) lattices do not fit very well; one needs a $\sqrt{19}\times \sqrt{19}R23.4^\mathrm{o}$ \MS\ supercell on top of a $4\times 4$ Ti(0001) supercell to get a mismatch below 1\%, see Table~\ref{tab:supercell}. The result therefore is a \MS/Ti(0001) interface that contains a substantial local strain, which explains why the binding energy is not very large, despite the bonding being chemisorption.   

The potential step $\Delta V$ at the \MS/Ti(0001) interface is negative, in contrast to the potential step at the \MS/Ag(111) interface for instance, which is positive, despite the fact that the work functions of Ti and Ag are very similar, see Table~\ref{tab:DeltaV}. We argued that physisorption should lead to a positive potential step because of the Pauli repulsion effect, and indeed \MS\ is physisorbed on Ag(111). Chemisorption, as in the case of \MS\ on Ti(0001), leads to a more drastic reorganization of the charge distribution at the interface, because of the formation of new chemical bonds. Upon the formation of these bonds there is apparently a net displacement of electronic density towards the sulfur atoms, which is not unreasonable as sulfur is more electronegative than Ti. This displacement results in an increase of the work function, i.e., a negative $\Delta V$.

\begin{figure}
\includegraphics[width=1.0\columnwidth]{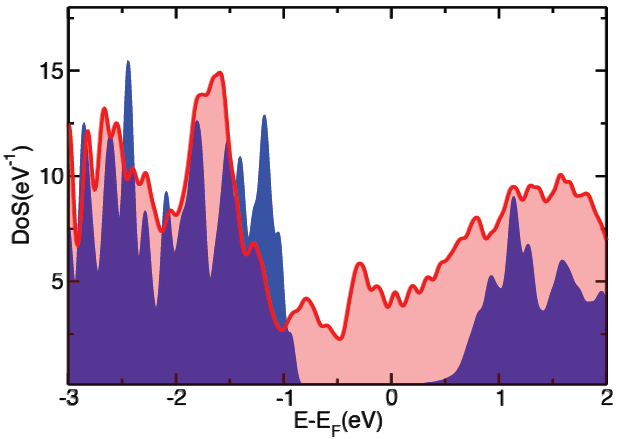}
\caption{(Color online) The red and the blue shaded areas indicate the DOS projected on the first \MS\ layer of \MS\ adsorbed on Ti(0001) and on the second \MS\ layer, respectively.}
 \label{fig:DOS_Ti}
\end{figure}

One expects that chemisorption also leads to a strong perturbation of the electronic structure of the adsorbate. Figure~\ref{fig:DOS_Ti} shows the density of states (DOS) of a \MS\ bilayer projected on the individual \MS\ layers. The DOS of the first (chemisorbed) layer is indeed strongly perturbed as compared to the DOS of a free-standing \MS\ layer. The \MS\ wave functions strongly hybridize with those of the underlying Ti substrate, and the resulting hybridized states give a non-zero DOS for energies all through the \MS\ band gap. It is sometimes argued that such interface states promote having a good (ohmic) \MS/metal contact.\cite{Popov:prl12,Kang:prx14} One could however also argue that chemisorption is harmful to obtaining a good contact, because it damages the integrity of the \MS\ layer. In Ref.~\onlinecite{HuiYuan:acs7} it is found that \MS/Ag gives a better contact than \MS/Ti, due to a much smoother interface in the former case, suggesting to prefer physisorption over chemisorption. 

It is not possible to define a Schottky barrier for a single \MS\ layer adsorbed on Ti(0001). Chemisorption affects the electronic structure of \MS\ to such an extend that its semiconducting character is lost. It is however possible to define a Schottky barrier for a second \MS\ layer that is adsorbed on the first layer. The first and second \MS\ layer are bonded by a van der Waals interaction, which does not perturb the electronic structure of the second layer significantly. Indeed the DOS of the second \MS\ layer is quite similar to that of a free-standing \MS\ layer with a clear band gap, see Fig.~\ref{fig:DOS_Ti}. The height of the Schottky barrier to the second layer is a sizable 0.54 eV. This mainly results from the fact that the first adsorbed \MS\ layer effectively increases the work function of the Ti substrate, see Table~\ref{tab:DeltaV}. The size of the Schottky barrier indicates that it is not possible to make an ohmic contact to pristine (undoped) \MS\ with Ti.\cite{Popov:prl12,Kang:prx14}


\section{Summary and Conclusions}
\label{sec:conclusions}
In this paper, we explore the adsorption of \MS\ on a range of metal substrates by means of first-principles DFT calculations. The (111) surfaces of Al, Ni, Cu, Pd, Ag, Pt and Au, and the (0001) surfaces of Mg, Co and Ti cover a range of metals with different interaction strengths, allowing for a systematic study of the metal-\MS\ interface. 

We compare the results obtained with different DFT functionals. In many cases the GGA/PBE density functional only captures a small part of the binding energy of \MS\ on a metal substrate, as compared to the opt88-vdW-DF van der Waals density functional, which indicates the importance of van der Waals interactions in the interface bonding. Nevertheless, the equilibrium binding distances obtained with both functionals are generally very similar, and so are the interface potential steps and Schottky barrier heights. Exceptions are cases for which van der Waals interactions essentially describe the whole bonding, where PBE fails completely. LDA tends to overbind, leading to shorter binding distances and larger interface potential steps.

The interface structure that results from adsorbing an \MS\ layer on a metal surface will be incommensurable in most cases, as the two lattices have a mismatch. We investigate the effects of the artificial strain introduced by approximating the lattice using a commensurable supercell. We conclude that these effects are moderate provided the \MS\ lattice parameter is kept at its optimized value, and the metal lattice is strained. Large lattice mismatches should however be avoided, and straining the \MS\ lattice can lead to very unphysical results.\cite{Popov:prl12} 

Of the metal substrates studied Ti is the one on which \MS\ is clearly chemisorbed. Adsorption of \MS\ on Ti(0001) is accompanied by a clear structural deformation of the Ti surface and of the \MS\ overlayer, due to the formation of bonds between the surface Ti atoms and the sulfur atoms at the interface. Formation of these interface bonds significantly alters the electronic structure of the \MS\ adsorbate. In particular, the interface states fill up the band gap of \MS, which makes defining a Schottky barrier for this layer meaningless. However, for a second, unperturbed, adsorbed \MS\ layer a Schottky barrier of 0.54 eV can be extracted. 

\MS\ is physisorbed on Au(111), where the bonding is almost completely due to van der Waals interactions, and the structure and electronic structure of \MS\ are hardly perturbed by the adsorption. The properties of \MS\ adsorbed on other metal substrates fall in the range between the two extreme cases (Ti and Au ), without the possibility of drawing a clear dividing line, as has been done for the adsorption of graphene or \BN\ on metal substrates.\cite{Giovannetti:prl08,Khomyakov:prb11,Bokdam:prb14a,Bokdam:prb14b}  

Experiments have focused foremost on Schottky barrier heights. Transport measurements on multilayer \MS\ devices generally yield small numbers for the Schottky barrier heights, i.e. 0.03-0.2 eV, for different metals,\cite{Qiu:apl12,Das:nanol13,Liu:acsnano12,Chen:nanol13,Kaushik:apl14}  whereas photoemission, photoconduction, and scanning tunneling spectroscopy give higher values 0.2-0.9 eV.\cite{Lince:prb87,Maurel:ss05,Fontana:scirep13,McDonnell:acsnano14} It has been suggested that the \MS\ samples used in devices is quite defective and inhomogeneous, such that the position of the Fermi level does not reflect an intrinsic property of \MS\ or of the \MS/metal contact,\cite{McDonnell:acsnano14,Yankowitz:nanol14} which obstructs a comparison to calculated results.

Our results for the 4d and 5d metals Ag, Au, Pd, and Pt agree qualitatively with those reported in previous calculations,\cite{Chen:nanol13,Gong:nanol14,Kang:prx14,Li:acs23} provided the \MS\ lattice is not stretched.\cite{Popov:prl12} Quantitatively, the reported Schottky barrier heights for these metals are $\sim 0.3$ eV larger than our results. These calculations were based upon the LDA functional, which tends to overbind, and to overestimate the metal work functions.\cite{Khomyakov:prb11,Bokdam:prb14a} Compressing the metal lattice, which is sometimes required to accommodate a lattice mismatch in a small supercell, does not help either, as that gives an even higher work function.\cite{Gong:nanol14} The same is likely true for simple metals such as Al and In.\cite{Gong:nanol14,Kang:prx14} The strong interaction we find for Ti is also found in LDA calculations.\cite{Popov:prl12,Kang:prx14,Li:acs23} In those calculations the lattice mismatch used was large, however, which can alter the interface interactions.

The overall picture emerging from these calculations is that \MS\ interacts strongly with the early transition metals, where it is clearly chemisorbed. The interaction with the late transition metals is much weaker, where the 3d metals interact stronger than the 4d and 5d metals. \MS\ interacts rather weakly with the simple metals, but the interaction increases for very low work function metals. In all but the strongly chemisorbed case, van der Waals forces play an important role in the interface interactions.

In case the interface interaction is weak (physisorption) the interface potential step can be understood as resulting from Pauli repulsion, which effectively decreases the substrate work function. The Schottky barrier is then simply calculated from the modified work function.
Strong interaction (chemisorption) leads to the formation of bonds between the substrate metal atoms and the adsorbate sulfur atoms. It increases the substrate work function if the electronegativity of the adsorbate is higher than that of the metal. If the \MS\ layer is chemisorbed, its electronic structure is perturbed to an extend that a Schottky barrier cannot be defined. However, a second adsorbed \MS\ layer then shows the characteristics of a single unperturbed layer. 

\acknowledgments
We acknowledge Taher Amlaki and Deniz \c{C}ak$\i$r for fruitful discussions. This work is part of the research program of the Foundation for Fundamental Research on Matter (FOM), which is part of the Netherlands Organization for Scientific Research (NWO). The use of supercomputer facilities was sponsored by the Physical Sciences Division (EW) of NWO.

\end{document}